\documentstyle{spacekap}

\begin{opening}

\title{THE COMBINATION OF HIPPARCOS DATA\\
WITH GROUND-BASED ASTROMETRIC MEASUREMENTS}

\author{R. Wielen}
\author{H. Schwan}
\author{C. Dettbarn}
\author{H. Jahreiss}
\author{H. Lenhardt}
\institute{Astronomisches Rechen-Institut, M\"onchhofstra{\ss}e 12-14,
           D--69120 Heidelberg, Germany}
\date{}

\end{opening}

\runningtitle{HIPPARCOS DATA AND GROUND-BASED MEASUREMENTS}
\runningauthor{R.~WIELEN~ET~AL.}

\begin{document}

\setcounter{topnumber}{2}

\begin{abstract}
The combination of HIPPARCOS measurements with suitable ground-based
astrometric data improves significantly the accuracy of the proper motions of
bright stars. The comparison of both types of data allows us also to identify
and to eliminate, at least partially, cosmic errors in the
quasi-instantaneously measured HIPPARCOS data which are caused by undetected
astrometric binaries. We describe a simple averaging method for the combination
of two independent compilation catalogues. The combination of the basic FK5
with HIPPARCOS leads to the Sixth Catalogue of Fundamental Stars (FK6). The
accuracy of the FK6 proper motions is higher than that of HIPPARCOS by a
factor of about 2 in the single-star mode, and by a factor of more than 4 in
the long-term prediction mode which takes cosmic errors into account. We
present also the error budget for a combination of the Boss General Catalogue
(GC) with HIPPARCOS data. We point out problems with known binaries, and
identify an ensemble of `astrometrically excellent stars'.


\end{abstract}

\section{Introduction}

The HIPPARCOS astrometry satellite has provided very accurate positions, proper
motions, and parallaxes for about 118\,000 stars (ESA 1997). Many people may
therefore be inclined to think that, at least for the stars observed by
HIPPARCOS, the ground-based observations, collected over more than two
centuries, are of no use anymore. It is correct that the systematic accuracy of
the HIPPARCOS data is so high that the system represented by the HIPPARCOS
positions and proper motions (ICRS) is superseding the earlier system (FK5),
derived from ground-based measurements. However, the random (`individual')
accuracy of the HIPPARCOS data, especially of the proper motions, can
nevertheless be significantly improved and checked by ground-based astrometric
measurements. Why is that so\,? The main reasons are: (1) The lower accuracy of
older ground-based measurements of stellar positions is often compensated by
the large spread in the epochs of observations. This allows us to derive very
accurate proper motions from the ground-based data alone or from a combination
with HIPPARCOS results (Wielen 1988). (2) Due to undetected astrometric
binaries, the HIPPARCOS proper motions can deviate significantly from the more
typical proper motions of the center-of-mass (Wielen 1995b, 1997, Wielen et
al. 1997). HIPPARCOS proper motions are derived essentially as `instantaneous'
proper motions from data collected within about 3 years only. In contrast, the
ground-based observations cover often more than two centuries. The derived
proper motions are then already rather close to `time-averaged' or `mean'
proper motions. Hence the ground-based results allow often to identify and to
correct `cosmic errors' in the instantaneously measured HIPPARCOS proper
motions.

\section{A Method of Combination}

In this section, we shall describe a simple averaging method for combining the
HIPPARCOS data (ESA 1997) with that of a ground-based compilation catalogue,
such as the FK5 (Fricke et al. 1988) or the GC (Boss et al. 1937). The method
neglects some of the correlations between the HIPPARCOS results for a given
star, namely the correlations between the results for $\alpha$ and $\delta$
and with the parallax. The advantage of the simple averaging method is its
transparency which allows easily to grasp the basic advantages and problems in
combining two (or more) catalogues. The results for the error estimates have
been already presented earlier (Wielen 1988).

Our method assumes that the two catalogues to be combined are based on
completely independ measurements. This is always fulfilled in combining
ground-based measurements with HIPPARCOS data.

Before combining the two catalogues, we have to reduce them to a common system
of positions and proper motions. In practice this means at present that we have
to reduce the ground-based catalogues to the HIPPARCOS system. This can be
done, for example, by using methods such as adopted in the construction of the
FK5 (Bien et al. 1977).

We call the position in one coordinate (say in declination $\delta$) $x$, the
corresponding proper motion $\mu$, and the central epoch $T$. The epoch $T$ is
chosen such that $x$ and $\mu$ are not correlated. The central epochs $T$ are
usually given explicitly in ground-based catalogues. For HIPPARCOS, the
individual epoch $T$ (and the corresponding position $x(T)$) can be easily
derived from the data given in the HIPPARCOS Catalogue (see e.g. Wielen 1997,
p. 374). The mean measuring errors of $x$ and $\mu$ are called $\varepsilon_x$
and $\varepsilon_\mu$. For systematically corrected ground-based catalogues,
these mean errors have to include the uncertainty of the applied systematic
corrections (e.g. Wielen et al. 1998). For HIPPARCOS we use the index $H$, for
the ground-based catalogue the index $F$, and for the catalogue obtained by
combining the two catalogues the index $C$.

The main conceptional device for combining the two catalogues is the
introduction of a third proper motion $\mu_0$, which is based on the two
central positions (Wielen 1988, 1995a):
\begin{equation}
\mu_0 = \frac{x_H(T_H) - x_F(T_F)}{T_H - T_F} \hspace*{0.5cm} .\\
\end{equation}
The mean error of $\mu_0$ is given by
\begin{equation}
\varepsilon_{\mu, 0} = \frac{(\varepsilon^2_{x, F} + \varepsilon^2_{x,
H})^{1/2}}{T_F - T_H} \hspace*{0.5cm} .\\
\end{equation}
We introduce the following `weights' $w$:
\begin{equation}
w_{x, H} = 1/\varepsilon^2_{x, H} \hspace*{0.5cm} ,
\end{equation}
\begin{equation}
w_{x, F} = 1/\varepsilon^2_{x, F} \hspace*{0.5cm} ,
\end{equation}
\begin{equation}
w_{\mu, H} = 1/\varepsilon^2_{\mu, H} \hspace*{0.5cm} ,
\end{equation}
\begin{equation}
w_{\mu, F} = 1/\varepsilon^2_{\mu, F} \hspace*{0.5cm} ,
\end{equation}
\begin{equation}
w_{\mu, 0} = 1/\varepsilon^2_{\mu, 0} \hspace*{0.5cm} .\\
\end{equation}
It can be shown that the position $x_C$, the proper motion $\mu_C$, and the
central epoch $T_C$ of the star in the combined catalogue $C$ can be derived as
`weighted averages'. The proof is not given here. The result is based on the
reconstruction of the full normal equations for the two unknowns $x$ and $\mu$
from the catalogued data $x(T), \, \mu, \, T, \, \varepsilon_x, \,
\varepsilon_\mu$\,. The method is an extended version of those described by
Kopff et al. (1964) or by Eichhorn (1974, Section 3.2.2). We obtain for the
weights in the combined catalogue $C$:
\begin{equation}
w_{x, C} = w_{x, H} + w_{x, F} \hspace*{0.5cm} ,
\end{equation}
\begin{equation}
w_{\mu, C} = w_{\mu, H} + w_{\mu, F} + w_{\mu, 0} \hspace*{0.5cm} .\\
\end{equation}
The new central epoch $T_C$ is given by
\begin{equation}
T_C = (w_{x, H} \, T_H + w_{x, F} \, T_F)/w_{x, C} \hspace*{0.5cm} .\\
\end{equation}
The position $x_C$ at $T_C$ follows as
\begin{equation}
x_C (T_C) = (w_{x, H} \, x_H (T_H) + w_{x, F} \, x_F (T_F))/w_{x, C}
\hspace*{0.5cm} .\\
\end{equation}
The combined proper motion $\mu_C$ is the weighted average of the three proper
motions $\mu_H, \, \mu_F,$ and $\mu_0$:
\begin{equation}
\mu_C = (w_{\mu, H} \,\, \mu_H + w_{\mu, F} \,\, \mu_F + w_{\mu, 0} \,\,
\mu_0)/w_{\mu, C}
\hspace*{0.5cm} .\\
\end{equation}
The mean errors $\varepsilon_{x, C}$ and $\varepsilon_{\mu, C}$ of $x_C (T_C)$
and $\mu_C$ are derived from
\begin{equation}
\varepsilon_{x, C} = (1/w_{x, C})^{1/2} \hspace*{0.5cm} ,
\end{equation}
\begin{equation}
\varepsilon_{\mu, C} = (1/w_{\mu, C})^{1/2} \hspace*{0.5cm} .
\end{equation}
Hence we have now obtained all the desired quantities for the combined
catalogue $C$.

In real applications, such as the FK6 (see Section 3), the proper motion
$\mu_0$ is often very accurate and governs the combined proper motion $\mu_C$.
For the positions, the accuracy of the HIPPARCOS position is usually by an
order of magnitude higher than that of the ground-based position. This leads to
$T_C \sim T_H$ and $x_C (T_C) \sim x_H (T_H)$. Nevertheless, for the
predictions of positions at epochs which
are not near to $T_H$, the accuracy of $\mu_C$
governs the accuracy of the predicted position. Hence the improved accuracy of
$\mu_C$ with respect to $\mu_H$ is very important for such a prediction of
positions.

We should mention that our averaging method has also an interesting `inverse'
application: If we know the combined catalogue $C$ and one of the two
underlying catalogues, we can derive back the second underlying catalogue. We
have carried out this `decomposition of a combined catalogue' for the
fundamental catalogues FK5, FK4, and FK3. In these cases, the catalogue
containing the `old' observations, e.g. the FK3 in the case of FK4, is
published and hence easily available. In contrast, the summary of the `new'
observations (in our example those between the closing dates of FK3 and FK4)
is not documented and hence unavailable. One can use the resulting `pieces' of
the FK5 (say FK5-FK4, FK4-FK3, FK3-NFK, NFK) for testing the linearity of the
motion of a star or for deriving a non-linear motion (similar to the G
solutions of HIPPARCOS). Clearly it would be more accurate to use all the
hundreds of individual observational catalogues for this purpose. This is
actually our intention for the future, but the decomposition method gives quite
rapidly first approximate results.

In the actual combination of a ground-based catalogue with the HIPPARCOS data,
we do not use the simple averaging method described above, but a strict method.
This strict method uses essentially the normal equations of rank 5 (two
positions: $\alpha, \, \delta$; two proper motions: $\mu_\alpha, \mu_\delta$;
one parallax: p), which can be derived from the HIPPARCOS Catalogue and the
ground-based catalogue. The strict method allows for correlations between
all the astrometric parameters in a catalogue. At first, it may be astonishing
that the ground-based data can influence the resulting parallax. This is due to
the fact that the ground-based data improve directly the proper motion. The
improved proper motion leads then indirectly to an improved value of the
parallax from the HIPPARCOS data. In reality, the correlation between the
parallax and the other astrometric parameters is usually so small for the
HIPPARCOS data that the improvement of the parallax in the combined catalogue
with respect to that derived from HIPPARCOS alone is negligible in most cases.
If the star is not single but indeed an undetected astrometric binary, then
the combination method leads to a less accurate parallax. In such a case the
`instantaneous' proper motion and position at $T_H$, not their `mean' values,
have to be used in deriving the parallax from the satellite observations, as
it was correctly done in the HIPPARCOS data reduction procedure.

\vspace*{-0.40 cm}

\section{Cosmic Errors and Statistical Astrometry}

The method of combining two catalogues described above is strictly valid for
truly single stars only, because the method assumes that a star moves linearly
in time on a straight line in space. Hence we call the result of the direct use
of the combination method the `single-star mode'.

In reality, we can form an ensemble of `apparently single stars' only. Such an
ensemble is usually spoiled by undetected astrometric binaries. In such an
ensemble the quasi-instantaneously measured HIPPARCOS proper motions and
positions suffer from `cosmic errors' (Wielen 1995b, 1997). The effect of
these errors on the combination of two catalogues can be properly treated by
the methods of statistical astrometry. The amount of the cosmic errors in the
HIPPARCOS proper motions has been determined by a comparison with the FK5 and
checked partially with the GC (Wielen 1995b, Wielen et al. 1997, 1998). For
example, averaged over about 1200 FK5 stars, we obtain a cosmic error in
$\mu_H$ of $c_\mu = 2.13$ mas/year, and $c_x \sim 10$ mas in $x_H$. The cosmic
error $c_\mu$ can be calibrated as a function of the stellar distance $r$
(obtained usually from the HIPPARCOS parallaxes). The derived cosmic errors in
$\mu_H$ and $x_H$ are usually larger than the measuring errors
$\varepsilon_{\mu, H}$ and $\varepsilon_{x, H}$, and are therefore quite
significant.

The `best prediction' of stellar positions based on a combination of two
catalogues affected by cosmic errors can be derived in general by applying the
concepts of statistical astrometry (Wielen 1997). At present, we use some
simplifying assumptions, because we lack the observational data for determining
the set of correlation functions needed for a full treatment. We assume that
the ground-based catalogue (e.g. the FK5 or the GC) is a fully time-averaged,
mean catalogue, without any cosmic errors left. This assumption is certainly
not strictly valid, but it should give already quite reasonable results.

Using the cosmic errors in $\mu_H$ and $x_H$, we can derive two other types of
predictions for the positions of the stars: (1) The `long-term prediction
mode': This prediction should be used for epochs $T$ which differ from $T_H =
1991.25$ by more than about 10 years. We obtain the long-term average mode by
replacing $\varepsilon^2_{\mu, H}$ by $\varepsilon^2_{\mu, H} + c^2_\mu$ , and
$\varepsilon^2_{x, H}$ by $\varepsilon^2_{x, H} + c^2_x$ . Otherwise, the
method of combining the two catalogues remains the same. (2) The `short-term
prediction mode': The cosmic errors are now added to the measuring errors of
the `mean' catalogue, i.e. $\varepsilon^2_{\mu, F}$ is replaced by
$\varepsilon^2_{\mu, F} + c^2_\mu$ and $\varepsilon^2_{x, F}$ by
$\varepsilon^2_{x, F} + c^2_x$ , while $\varepsilon_{\mu, H}$ and
$\varepsilon_{x, H}$ are not changed. The short-term prediction is valid for
epochs which differ from $T_H$ by a few years only. For a smooth transition
from the short-term to the long-term prediction, we would have to know the
proper correlation functions (see Wielen 1997).

\begin{samepage}

\begin{table}[t]
\caption{
Error budget for FK6 proper motions in the `single-star mode'
}
\begin{center}
\begin{tabular}{lcc}
\hline\\[-1.0ex]
\multicolumn{3}{c}{Typical mean errors of proper motions}\\
\multicolumn{3}{c}{(in one component, averaged over $\mu_{\alpha\ast}$ and
$\mu_\delta$; units: mas/year)}\\[1.5ex]\hline\\[-2.5ex]
                              & rms average & median\\[0.5ex]\hline\\[-1.8ex]
HIPPARCOS                     & 0.82        & 0.63\\[1.5ex]
FK5\\
\hspace*{0.5cm}random         & 0.76        & 0.64\\
\hspace*{0.5cm}system         & 0.28        & 0.24\\
\hspace*{0.5cm}total          & 0.81        & 0.70\\[1.5ex]
$\mu_0$ (total)               & 0.58        & 0.49\\[1.5ex]
FK6                           & 0.35        & 0.33\\[1.5ex]
ratio of HIPPARCOS to FK6 errors \hspace*{1.00 cm} & 2.3\hphantom{0} &
                                    1.9\hphantom{0}\\[0.5ex]\hline
\end{tabular}
\end{center}
\end{table}

\begin{table}[t]
\caption{
Error budget for FK6 proper motions in the `long-term prediction mode'
}
\begin{center}
\begin{tabular}{lcc}\hline\\[-1.0ex]
\multicolumn{3}{c}{Typical mean errors of proper motions}\\
\multicolumn{3}{c}{(in one component, averaged over $\mu_{\alpha\ast}$ and
                     $\mu_\delta$; units: mas/year)}\\[1.5ex]\hline\\[-2.5ex]
                              & rms average & median\\[0.5ex]\hline\\[-1.8ex]
HIPPARCOS\\
\hspace*{0.5cm}measuring error & 0.68       & 0.60\\
\hspace*{0.5cm}cosmic error    & 2.13       & (2.04)\\
\hspace*{0.5cm}total           & 2.24       & (2.13)\\[1.5ex]
FK5\\
\hspace*{0.5cm}random          & 0.77       & 0.66\\
\hspace*{0.5cm}system          & 0.28       & 0.24\\
\hspace*{0.5cm}total           & 0.82       & 0.70\\[1.5ex]
$\mu_0$ (total)                & 0.59       & 0.51\\[1.5ex]
FK6                            & 0.49       & 0.45\\[1.5ex]
ratio of HIPPARCOS (total) to FK6 errors & 4.6\hphantom{0}  &
                                             (4.7)\hphantom{.}\\[0.5ex]\hline
\end{tabular}
\end{center}
\vspace*{-0.20 cm}
\end{table}

\end{samepage}

\section{The Sixth Catalogue of Fundamental Stars (FK6)}

The FK6 is derived by combining the ground-based data summarized in the FK5
(Part I, Fricke et al. 1988) with the HIPPARCOS measurements (ESA 1997). A more
detailed description of the construction of the FK6 is given in another paper
(Wielen et al. 1998).

In Table I, we give the formal error budget of the FK6 for all the basic
fundamental stars in the single-star mode. Table II presents the error budget
of the FK6 for 1202 stars in the long-term prediction mode. The results of
the short-term prediction mode differ only slightly from that of the direct use
of the HIPPARCOS data and are not given here. We should warn those readers
who would like to test our numerical results quoted for the FK6 by using the
other quantities of Tables I and II as input data. The method of combination
works star by star, while the tables give ensemble averages. Since
1/$<\varepsilon^2>$ is in general not exactly equal to $<1/\varepsilon^2>$,
such a test must slightly fail.

How much better are the FK6 proper motions than the direct HIPPARCOS proper
motions\,? The gain in accuracy in the single-star mode is about a factor of 2,
and in the long-term prediction mode a factor of more than 4. These
improvements are certainly significant and favour strongly the use of the FK6.

\begin{samepage}

\begin{table}[t]
\caption
{
Error budget for GC+HIP proper motions in the `single-star mode'
}
\begin{center}
\begin{tabular}{lcccccc}
\hline\\[-1.0ex]
\multicolumn{7}{c}{Typical mean errors of proper motions}\\
\multicolumn{7}{c}{(in one component, averaged over $\mu_{\alpha\ast}$ and
$\mu_\delta$; units: mas/year)}\\[1.5ex]\hline\\[-2.5ex]
Sample of stars: & \multicolumn{2}{c}{29\,717 GC} &
\multicolumn{2}{c}{11\,773 GC} &
\multicolumn{2}{c}{1534 FK from GC}\\
& rms av. & median & rms av. & median & rms av. &
median\\[0.5ex]\hline\\[-1.8ex]
HIPPARCOS  & 1.47  & 0.72 & 0.76 & 0.68 & 0.82 & 0.63\\[1.5ex]
GC (total) & 10.58\hphantom{1} & 9.29 & 8.60 & 7.45 & 3.36 & 2.39\\[1.5ex]
$\mu_0$\,(total) & 1.79 & 1.73 & 1.43 & 1.44 & 0.66 & 0.55\\[4ex]
GC+HIP & 0.73 & 0.63 & 0.62 & 0.57 & 0.42 & 0.39\\[4ex]
ratio of HIPPARCOS & 2.0\hphantom{-} & 1.14 & 1.23 & 1.19 & 2.0\hphantom{-} &
1.6\hphantom{-}\\ to GC+HIP errors\\[0.5ex]\hline
\end{tabular}
\end{center}
\vspace*{-0.20 cm}
\end{table}

\begin{table}[t]
\caption
{
Error budget for GC+HIP proper motions in the `long-term prediction mode'
}
\begin{center}
\begin{tabular}{lcccc}
\hline\\[-1.0ex]
\multicolumn{5}{c}{Typical mean errors of proper motions}\\
\multicolumn{5}{c}{(in one component, averaged over $\mu_{\alpha\ast}$ and
$\mu_\delta$; units: mas/year)}\\[1.5ex]\hline\\[-2.5ex]
Sample of stars: & \multicolumn{2}{c}{11\,773 GC} &
\multicolumn{2}{c}{1201 FK from GC}\\
& rms average & median & rms average & median\\[0.5ex]\hline\\[-1.8ex]
HIPPARCOS\\
\hspace*{0.5cm} measuring error & 0.76 & 0.68 & 0.68 & 0.60\\
\hspace*{0.5cm} cosmic error & 1.80 & (1.66) & 2.13 & (2.04)\\
\hspace*{0.5cm} total & 1.95 & (1.79) & 2.24 & (2.13)\\[1.5ex]
GC (total) & 8.60 & 7.45 & 3.47 & 2.47\\[1.5ex]
$\mu_0$\,(total) & 1.43 & 1.44 & 0.67 & 0.57\\[4ex]
GC+HIP & 1.07 & 1.05 & 0.61 & 0.55\\[4ex]
ratio of HIPPARCOS (total) & 1.8\hphantom{.} & (1.7)\hphantom{.} &
3.7\hphantom{1} & (3.9)\hphantom{-}\\
to GC+HIP errors\\[0.5ex]\hline
\end{tabular}
\end{center}
\vspace*{-0.20 cm}
\end{table}

\end{samepage}

\section{Combination of the GC with HIPPARCOS data}

For the combination with the very accurate HIPPARCOS data, old ground-based
observations are especially valuable, because they have large epoch differences
with respect to $T_H$ and often still a reasonable accuracy. A carefully
compiled catalogue of old observations of more than 30\,000 stars is the GC
(Boss et al. 1937). Therefore we have applied our combination method also to
the GC. We call the resulting combined catalogue GC+HIP.

In Table III we present the formal error budget of the GC+HIP in the
single-star mode. Table IV gives the error budget of the GC+HIP in the
long-term prediction mode. For the cosmic errors $c_\mu$ and $c_x$ in the
HIPPARCOS proper motions and positions of the GC stars, we use the results
determined from the basic FK5 stars. This is certainly an approximation only:
The GC stars are in general fainter in apparent magnitude than the basic FK5
stars (by about 1.9 magnitudes), and they are more distant (median parallax of
GC stars: 6.8 mas, of basic FK5 stars: 11.3 mas). Nevertheless, the comparison
of $\mu_0$ of the GC+HIP with $\mu_H$ for GC stars has essentially lead to the
same values of $c_\mu$ and $c_x$ as for the basic FK5 stars (Wielen et al.
1998).

How important the old observations are, is impressively shown by the
combination of the GC data with the HIPPARCOS measurements for 1534 basic FK
stars. From a comparison of the last columns of Table III with Table I we see
that for these FK stars the combined catalogue GC+HIP is only slightly less
accurate than the FK6 (= FK5+HIP) itself.

The first columns of Table III give the formal error budget for all the 29\,717
GC stars which we have identified in the HIPPARCOS Catalogue. In the columns in
the middle of Table III, we give the results for 11\,773 `apparently single' GC
stars. This ensemble contains only those GC stars which have no special
HIPPARCOS solutions in the DMSA, are no `suspected' binaries according to
HIPPARCOS, do not occur in the CCDM, and have mean errors of $\mu_{0, \alpha
\ast}$ and $\mu_{0, \delta}$ smaller than 2 mas/year.

In the single-star mode (Table III), the rms average for all the GC stars
indicates a significant gain by a factor of 2 with respect to HIPPARCOS, while
the median shows a smaller improvement (by 14 percent) for these stars. For the
11\,773 GC stars (apparently single and with an accurate $\mu_0$) the gain is
about 20 percent. The gain is much higher in the long-term prediction mode
(Table IV). For the 11\,773 GC stars, the accuracy in the GC+HIP proper motion
is by nearly a factor of 2 higher than for the HIPPARCOS proper motion (if we
now include the cosmic error $c_\mu$ of $\mu_H$).

In individual comparison of $\mu_0$ derived from the GC and HIPPARCOS with a
typical accuracy of 2 mas/year, with $\mu_H$ with a measuring error of about 1
mas/year, allows us to identify GC stars with large cosmic errors (say above 6
mas/year) in their instantaneously measured HIPPARCOS proper motions. These
stars are good candidates for being checked by other observational techniques
for their probable binary nature.

\section{Known Binaries and Astrometrically Excellent Stars}

Known binaries require usually a rather individual treatment in the process of
combining ground-based data with the HIPPARCOS measurements. The main problem
is to find a common `reference point' for using the method of combination
described in Section 2. Such reference points may be the center-of-mass, an
instantaneous photo-center, a time-averaged photo-center, the separated
components A and B of a double star, etc. Often we have to use statistical
methods for `harmonizing' the reference points of the ground-based data and
of the satellite measurements. A more detailed description of the problems with
double stars is given by Wielen et al. (1998).

Even after doing the best you can do, the accuracy of the position and of the
proper motion of a binary is often lower by an order of magnitude or more
compared to single stars. In order to help the user who is interested in stars
of highest astrometric accuracy, we are flagging in the FK6 and in the GC+HIP
those stars which behave like well-measured single stars. We call this sample
`astrometrically excellent stars'. These stars should not show any known
disturbing duplicity. The measuring accuracy of their positions and proper
motions should be good. But most important, their proper motions $\mu_H$ and
$\mu_0$, and, in the case of the FK6, $\mu_{FK5}$ should agree within the
measuring errors. This agreement indicates, at least statistically, that the
individual cosmic error of $\mu_H$ is rather small for these stars.

\section{Conclusions}

The combination of HIPPARCOS measurements with ground-based astrometric data
allows us
to improve the accuracy of the proper motions of thousands of stars.
Improved data for the basic fundamental stars are presented in the FK6. For
many other stars for which the old observations are summarized in the GC,
improved data have also been derived. The ground-based data are especially
valuable for identifying large individual cosmic errors in the
quasi-instantaneously measured HIPPARCOS proper motions of undetected binaries.


\begin{thebibliography}{}
\bibitem[]{}
Bien, R., Fricke, W., Lederle, T., Schwan, H.: 1977, {\it Ver\"off. Astron.
Rechen-Inst. Heidelberg} No. 27

\bibitem[]{}
Boss, B., Albrecht, S., Jenkins, H., Raymond, H., Roy, A.J., Varnum,
W.B., Wilson, R.E.: 1937, {\it General Catalogue of 33\,342 Stars for the Epoch
1950}, Carnegie Institution of Washington, Publ. No. 486

\bibitem[]{}
Eichhorn, H.: 1974, {\it Astronomy of Star Positions}, Frederick Ungar Publ.
Co., New York

\bibitem[]{}
ESA: 1997, {\it The Hipparcos Catalogue}, ESA SP-1200

\bibitem[]{}
Fricke, W., Schwan, H., Lederle, T., Bastian, U., Bien, R., Burkhardt, G., du
Mont, B., Hering, R., J\"ahrling, R., Jahrei{\ss}, H., R\"oser, S.,
Schwerdtfeger, H.M., Walter, H.G.: 1988, {\it Ver\"off. Astron. Rechen-Inst.
Heidelberg} No. 32

\bibitem[]{}
Kopff, A., Nowacki, H., Strobel, W.: 1964, {\it Ver\"off. Astron. Rechen-Inst.
Heidelberg} No. 14

\bibitem[]{}
Wielen, R.: 1988, in: IAU Symposium No. 133, {\it Mapping the Sky}, eds. S.
Debarbat, J.A. Eddy, H.K. Eichhorn, A.R. Upgren, Kluwer Publ. Comp., Dordrecht,
p. 239

\bibitem[]{}
Wielen, R.: 1995a, {\it Astron. Astrophys.} {\bf 302}, 613

\bibitem[]{}
Wielen, R.: 1995b, in: {\it Future Possibilities for Astrometry in Space}, eds.
M.A.C. Perryman, F. van Leeuwen, ESA SP-379, p. 65

\bibitem[]{}
Wielen, R.: 1997, {\it Astron. Astrophys.} {\bf 325}, 367

\bibitem[]{}
Wielen, R., Schwan, H., Dettbarn, C., Jahrei{\ss}, H., Lenhardt, H.: 1997, in:
{\it Hipparcos Venice '97}, Presentation of the Hipparcos and Tycho Catalogues
and first astrophysical results of the Hipparcos space astrometry mission, eds.
Battrick, B., Perryman, M.A.C., Bernacca, P.L., ESA SP-402, p. 727

\bibitem[]{}
Wielen, R., Schwan, H., Dettbarn, C., Jahrei{\ss}, H., Lenhardt, H.: 1998, in:
{\it The Message of the Angles. Astrometry 1798\,-\,1998}. Proceedings of the
International Spring Meeting of the Astronomische Gesellschaft, held in Gotha,
11-15 May 1998, eds. P. Brosche, W.R. Dick, O. Schwarz, R. Wielen (in press)
\end{thebibliography}
\end{document}